\documentclass[aps,prb,twocolumn,groupedaddress]{revtex4}
\usepackage{graphicx}
\usepackage{color}

\begin{document}


\title{
Enhancement of electron correlation due to the molecular dimerization 
in organic superconductors $\beta$-(BDA-TTP)$_{2}X$ ($X$=I$_3$, SbF$_6$)
}



\author{Hirohito Aizawa}
\email[Electronic address: ]{aizawa@kanagawa-u.ac.jp}
\affiliation{
 Institute of Physics, 
 Kanagawa University, 
 Yokohama, Kanagawa 221-8686, Japan}

\author{Kazuhiko Kuroki}
\affiliation{
 Department of Physics, 
 Osaka University, 
 Toyonaka, Osaka 560-8531, Japan}

\author{Jun-ichi Yamada}
\affiliation{
 Department of Material Science, 
 University of Hyogo, 
 Ako-gun, Hyogo 678-1297, Japan}


\date{\today}

\begin{abstract}
%
We perform a first principles band calculation for 
quasi-two-dimensional organic superconductors 
$\beta$-(BDA-TTP)$_{2}$I$_{3}$ and $\beta$-(BDA-TTP)$_{2}$SbF$_{6}$. 
The first principles band structures between the I$_{3}$ and SbF$_{6}$ salts 
are apparently different. 
We construct a tight-binding model for each material 
which accurately reproduces the first principles band structure. 
The obtained transfer energies give the differences such as 
(i) larger dimerization in the I$_{3}$ salt than the SbF$_{6}$ salt, 
and (ii) different signs and directions of the inter-stacking transfer energies. 
To decompose the origin of the difference into the dimerization and 
the inter-stacking transfer energies, 
we adopt a simplified model by eliminating the dimerization effect 
and extract the difference caused by the inter-stacking transfer energies. 
From the analysis using the simplified model, 
we find that the difference of the band structure 
comes mainly from the strength of dimerization. 
To compare the strength of the electron correlation 
having roots in the band structure, 
we calculate the physical properties originated from 
the effect of the electron correlation such as the spin susceptibility 
applying two particle self-consistent (TPSC) method. 
We find that the maximum value of the spin susceptibility of the I$_{3}$ salt 
is larger than that of the SbF$_{6}$ salt. 
Hypothetically decreasing the dimerization within the model of the I$_{3}$ salt, 
the spin susceptibility takes almost the same value as that of the SbF$_6$ salt for the same magnitude of the dimerization. 
We expect that the different ground state 
between the I$_{3}$ and SbF$_{6}$ salt 
mainly comes from the strength of the dimerization 
which is apparently masked in the band calculation 
along a particular $k$-path. 
\end{abstract}


\maketitle


\section{Introduction}
\label{Introduction} 

\begin{figure}[!htb]
 \centering
  \includegraphics[width=8.0cm]{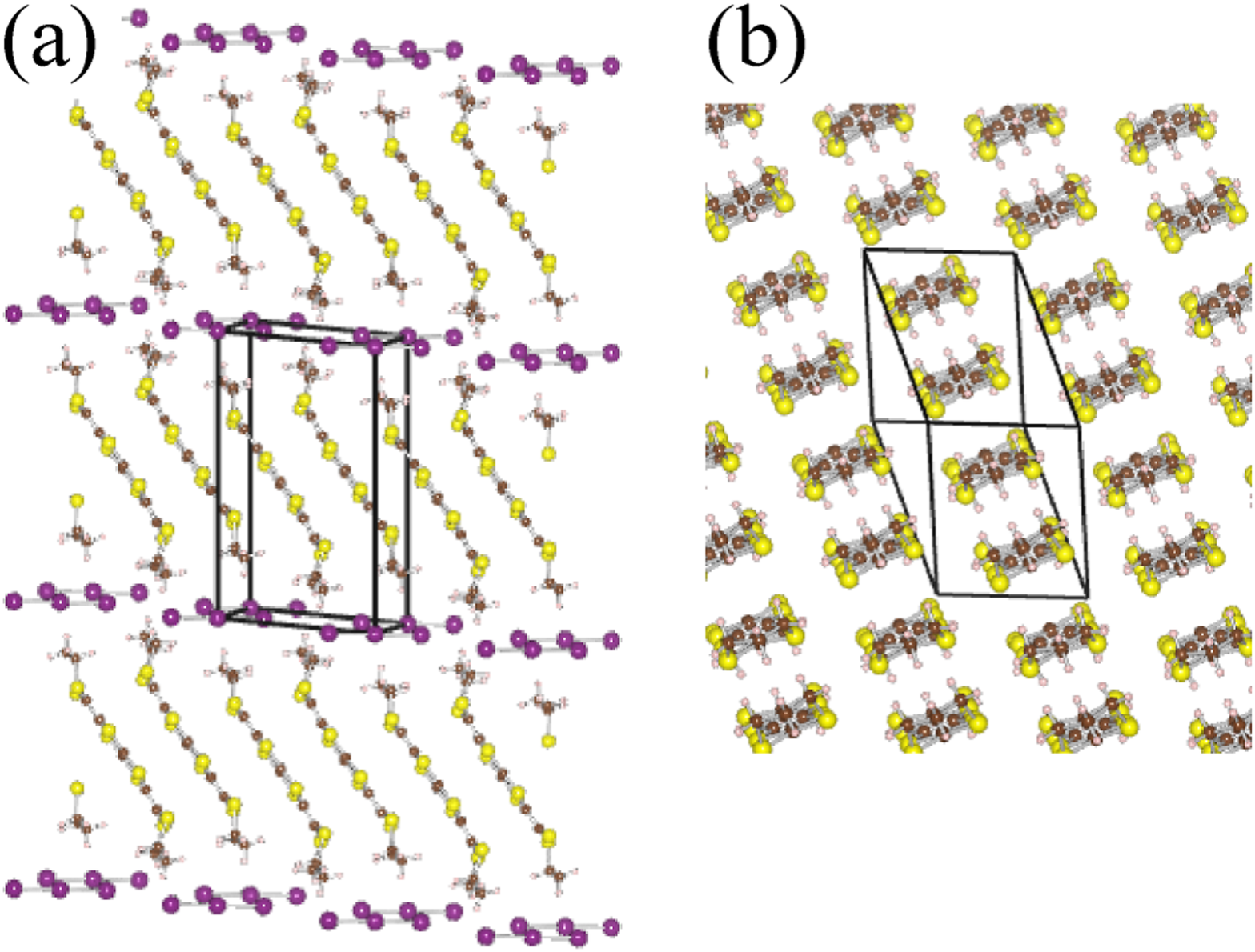}
 \caption{ (color online) 
 Crystal structure of the I$_3$ salt from 
 (a) the side view and 
 (b) the conductive layer of the BDA-TTP molecules. 
 }
 \label{fig1}
\end{figure}

There have been attempts to synthesize 
strongly correlated electron systems in organic conductors 
by applying chemical modification to stable metallic donor molecules. 
For example, there are ($S$,$S$)-DMBEDT-TTF 
\cite{Zambounis-Carl-am-4-33} 
and \textit{meso}-DMBEDT-TTF 
\cite{Kimura-Maejima-cc--2454}, 
where two methyl groups are attached to BEDT-TTF, 
and they are pressure-induced superconductors. 
In the present article, 
we theoretically study superconductors based on BDA-TTP molecule, 
which is extended to six-membered-ring from five-membered-ring 
in the $\sigma$-bond framework of BDH-TTP molecule 
\cite{Yamada-Watanabe-JACS-104-5057}. 
The actual materials 
are $\beta$-(BDA-TTP)$_2$I$_3$ and $\beta$-(BDA-TTP)$_2$SbF$_6$, 
which will be abbreviated as I$_3$ and SbF$_6$ salts, respectively.
In both materials, conductive layer is the BDA-TTP layer, 
and the anion layer separates the adjacent conductive layers 
as shown in Fig. \ref{fig1} (a). 
Molecular configuration in the conductive layer is the $\beta$-type 
as shown in Fig. \ref{fig1} (b). 
Both materials consist of the stacking structure of the BDA-TTP molecules. 
However, they are somewhat different in that the inter-stacking direction 
is slightly tilted in the I$_3$ salt, 
but almost side-by-side for the SbF$_6$ salt, 
which will be shown later. 

The I$_3$ salt is an insulator at ambient pressure, 
and the superconductivity appears around 10 K 
under hydrostatic pressure of above 10 kbar 
\cite{Yamada-Fujimoto-CC-2006-1331}. 
Recently, applying uniaxial strain along the $c$-axis 
has given higher $T_{\rm c}$ 
\cite{Kikuchi-Isono-JACS-133-19590}. 
Applying the uniaxial compression once increases the $T_c$ and 
takes a maximum before it decreases 
\cite{Ito-Ishihara-PRB-78-172506}. 
It is considered that 
applying the pressure in the I$_3$ salt 
increases the overlap between the upper and lower bands, 
which gradually changes the character of the system from 
a strongly correlated half-filled system 
to a moderately correlated quarter filled system. 
The $c$-axis strain more efficiently increases the band-width of the overlap. 
As the electron correlation is reduced to some extent by pressure, 
the insulating nature of the material is lost, and superconductivity appears 
\cite{Kikuchi-Isono-JACS-133-19590}. 
Theoretically, 
Nonoyama \textit{et al.} have studied the nature of the charge ordering state 
and the pairing mechanisms in the model of the I$_3$ salt 
derived from the extended H$\ddot{\rm u}$ckel band structure
\cite{Nonoyama-Maekawa-JPCS-132-012013}. 

The SbF$_6$ salt exhibits superconductivity 
at 7.5 K at ambient pressure 
\cite{Yamada-Watanabe-JACS-123-4174}. 
As for the SbF$_6$ salt, there have been some controversies regarding 
both the anisotropy of the Fermi surface 
\cite{Choi-Jobilong-PRB-67-174511,Yasuzuka-Koga-JPSJ-81-035006}
and the directions of the nodes in the superconducting gap
\cite{Shimojo-Ishiguro-JPSJ-71-717,Tanatar-Ishiguro-PRB-71-024531,Nomura-Muraoka-PhysicaB-404-562,Yasuzuka-Koga-ICSM2010-5Ax-10}. 
In our previous study \cite{Aizawa-Kuroki-NJP-14-113045} 
for the $\beta$-(BDA-TTP)$_{2}M$F$_6$ ($M$=P, As, Sb and Ta), 
we have obtained the band structure from the first principles band calculation, 
and suggested the origin of the differences from 
the extended H$\ddot{\rm u}$ckel band structure\cite{huckel-comment}. 
Also, there have been some studies on pairing mechanisms   
mediated by spin and/or charge fluctuations 
in the model of $\beta$-(BDA-TTP)$_{2}X$. 
As for the $M$F$_6$ ($M$= As, Sb) salts, 
adopting models derived from the extended H$\ddot{\rm u}$ckel calculation, 
Nonoyama {\it et al.} 
\cite{Nonoyama-Maekawa-JPSJ-77-094703}
have applied random phase approximation (RPA) to the two band model, 
while Suzuki {\it et al.} 
\cite{Suzuki-Onari-JPSJ-80-094704} 
have applied the fluctuation exchange (FLEX) approximation 
to the original two-band model and the single-band dimer model. 
Recently, we have constructed the tight-binding model derived from 
the first principles band calculation, 
studied the pairing symmetry of the gap function within the 
spin fluctuation mediated pairing
\cite{Aizawa-Kuroki-NJP-14-113045}. 

In the present study, 
given the difference in the ground state 
between the I$_3$ salt and the SbF$_6$ salt, 
we focus on the difference in the electronic structure between the two salts. 
In fact, despite the similar lattice structure, 
the band structure of the I$_3$ salt 
\cite{Yamada-Fujimoto-CC-2006-1331}
and that of the SbF$_6$ salt 
\cite{Yamada-Watanabe-JACS-123-4174} 
obtained by the extended H$\ddot{\rm u}$ckel method 
are known to be very different. 
Here, we perform the first principles band calculation 
for $\beta$-(BDA-TTP)$_2$I$_3$ 
and construct an effective tight-binding model 
that reproduces the first principles band structure. 
We compare the band structure of the I$_3$ salt 
to that of the SbF$_6$ salt obtained 
in our previous study\cite{Aizawa-Kuroki-NJP-14-113045}, 
and pin down the origin of the apparently large differences. 
In particular, we study the relation between the strength of 
the electron correlation and the molecular dimerization.  
We consider the Hubbard model by introducing repulsive interaction between 
the electrons on the same BDA-TTP molecule.
Then, we study the effect of the electron correlation 
by applying the two particle self-consistent (TPSC) method, 
and present
quantities such as the spin susceptibility 
against the temperature and dimerization strength,   
which reflect physical properties originating from the electron correlation. 
We conclude that the ground state of the I$_{3}$ salt 
differs from that of the SbF$_{6}$ salt 
due to the strength of the dimerization.

\section{Method} 
\label{Method} 

\subsection{first principles band calculation and model construction} 
\label{First principles band calculation}

We perform first principles band calculation 
using all-electron full potential linearized augmented plane-wave (LAPW) 
+ local orbitals (lo) method within the framework of WIEN2k 
\cite{WIEN2K}. 
This implements the density functional theory (DFT) 
with different possible approximation 
for the exchange correlation potentials. 
The exchange correlation potential is calculated 
using the generalized gradient approximation (GGA). 

The single-particle wave functions in the interstitial region are 
expanded by plane waves with a cut-off of $R_{\rm MT} K_{\rm max}=3.0$
due to the presence of the hydrogen atom, 
where $R_{\rm MT}$ denotes the smallest muffin-tin radius 
and $K_{\rm max}$ is the maximum value of $K$ vector 
in the plane wave expansion. 
In the I$_{3}$ salt, 
the muffin-tin radii are assumed to be 
2.50, 1.62, 1.15, and 0.62 atomic units (a.u.) 
for I, S, C, and H, respectively. 
$K_{\rm max}$ is taken as 4.8, 
and the plane wave cutoff energy is 318.6 eV. 
In the SbF$_{6}$ salt, 
the muffin-tin radii are assumed to be 
1.74, 1.74, 1.62, 0.83, and 0.45 a.u. 
for Sb, F, S, C, and H, respectively. 
$K_{\rm max}$ is taken as 6.7, 
and the plane wave cutoff energy is 604.7 eV. 
Calculations were performed using 
6$\times$3$\times$9 $k$-points for the I$_{3}$ salt and 
7$\times$3$\times$9 $k$-points for the SbF$_{6}$ salt 
in the irreducible Brillouin zone. 
We adopt the lattice structure determined experimentally 
for each materials 
\cite{Yamada-Fujimoto-CC-2006-1331,Yamada-Watanabe-JACS-123-4174}, 
and we do not relax the atomic positions in the calculation.

Having done the first principles band calculation, 
we then construct a tight-binding model which accurately 
reproduces the first principles band structure. 
From the lattice structure of the two materials, 
we regard one molecule as a site 
and consider a two-band (two sites per unit cell) tight-binding model 
to fit the first principles band structure. 
The tight-binding Hamiltonian, $H_{0}$, is written in the form 
\begin{eqnarray}
 H_{0}=\sum_{\left< i \alpha: j \beta \right>, \sigma}
  \left\{ t_{i \alpha: j \beta} 
   c_{i \alpha \sigma}^{\dagger} c_{j \beta \sigma} + {\rm H. c.} 
  \right\}, 
\label{H0ij}
\end{eqnarray} 
where 
$i$ and $j$ are unit cell indices, 
$\alpha$ and $\beta$ specifies the sites in a unit cell, 
$c_{i \alpha \sigma}^{\dagger}$ ($c_{i \alpha \sigma}$ ) is 
a creation (annihilation) operator with spin $\sigma$ 
at site $\alpha$ in the $i$-th unit cell, 
$t_{i \alpha: j \beta}$ is the electron transfer energy 
between $(i, \alpha)$ site and $(j, \beta)$ site, and 
$\left< i \alpha: j \beta \right>$ represents 
the summation over the bonds corresponding to the transfer.

By Fourier transformation, eq. (\ref{H0ij}) is rewritten as 
\begin{eqnarray}
 H_{0}=\sum_{\textbf{\textit k}, \sigma, \alpha, \beta}
  \varepsilon_{\alpha \beta}\left( \textbf{\textit{k}} \right)
   c_{\textbf{\textit k} \alpha \sigma}^{\dagger} 
   c_{\textbf{\textit k} \beta \sigma}, 
\end{eqnarray}
where $\varepsilon_{\alpha \beta}\left( \textbf{\textit{k}} \right)$ is 
the site-indexed kinetic energy represented in $\textbf{\textit{k}}$-space. 
The band dispersion is given by diagonalizing 
the matrix $\varepsilon_{\alpha \beta}\left( \textbf{\textit{k}} \right)$, 
\begin{eqnarray}
 \varepsilon_{\alpha \beta}\left( \textbf{\textit{k}} \right)
  =\sum_{\gamma}
  d_{\alpha \gamma}\left( \textbf{\textit{k}} \right)
  d_{\beta \gamma}^{*}\left( \textbf{\textit{k}} \right) 
  \xi_{\gamma}\left( \textbf{\textit{k}} \right), 
  \label{eps-ab}
\end{eqnarray}
where $\xi_{\gamma}\left( \textbf{\textit{k}} \right)$ gives 
the band dispersion of the $\gamma$-th band 
measured from the chemical potential, 
and $d_{\alpha \gamma}\left( \textbf{\textit{k}} \right)$ 
is the unitary matrix that gives the unitary transformation.

We adopt the two-band Hubbard model 
obtained by adding the on-site (intra-molecule) repulsive interaction 
to the tight-binding model derived from 
the fitting of the first principles band structure. 
The Hubbard Hamiltonian, $H$, is 
\begin{eqnarray}
 H=H_{0}
  +\sum_{i \alpha} U_{0}
  n_{i \alpha \uparrow} n_{i \alpha \downarrow}
  \label{Hij}
\end{eqnarray} 
where $U_{0}$ is the bare on-site interaction and $n_{i \alpha \sigma}$ 
is the number operator of the electron 
on the $\alpha$-site in the $i$-th unit cell. 
Since both salts are configured as a form of $D_2 X$ 
where $D$ is the donor molecule and $X^{-1}$ is the anion, 
the band-filling is  $1/4$-filled in the hole representation 
($3/4$-filled in the electron representation).

\subsection{Two particle self consistent method} 
\label{Two particle self consistent method}

To deal with the electron correlation effect 
arising from the on-site repulsion, 
we apply TPSC to the multi-site Hubbard model 
given by eq. (\ref{Hij}) as follows. 
The bare susceptibility in the site-representation is given by 
\begin{eqnarray}
 \chi^{0}_{\alpha \beta} \left( q \right)
 &=&-\frac{T}{N_c} \sum_{ k }
  G^{0}_{\alpha \beta }\left( k+q \right) G^{0}_{\beta  \alpha}\left( k \right),   
 \label{chi0}
\end{eqnarray}
where $T$ and $N_c$ are the temperature and the total number of unit cells, 
respectively, 
and $G^{0}_{\alpha \beta}\left( k \right)$ is the bare Green's function
given as 
\begin{eqnarray}
 G^{0}_{\alpha \beta} \left( k \right)
 &=& \sum_{\gamma}
 d_{\alpha \gamma}\left( \textbf{\textit{k}} \right) 
 d_{\beta \gamma}^{*}\left( \textbf{\textit{k}} \right) 
 \frac{1}{ i \varepsilon_{n}-\xi_{\gamma} \left( \textbf{\textit{k}} \right)}. 
 \label{g0}
\end{eqnarray}
Here, we introduce the abbreviations 
$k=\left( \textbf{\textit{k}}, i \varepsilon_{n} \right)$ 
and 
$q=\left( \textbf{\textit{q}}, i \omega_{m} \right)$ 
for the fermionic and bosonic Matsubara frequencies.  
The indices $\alpha\beta$ means ($\alpha$ $\beta$)-element of 
the matrix such as $\hat{\chi}^{0}\left( q \right)$.

TPSC has been applied to single-site systems
\cite{Vilk-Tremblay-JPIF-7-1309,Otuski-PRB-85-104513}, 
multi-site system, 
\cite{Arya-Sriluckshmy-arXiv-1504-06373}
and multi-orbital system 
\cite{Miyahara-Arita-PRB-87-045113}. 
By applying TPSC, we can consider the local vertex correction 
in both spin and charge channels within a self-consistent procedure. 
In the TPSC, using the bare susceptibility given by eq. (\ref{chi0}), 
the spin and charge susceptibilities are obtained as 
\begin{eqnarray}
 \hat{\chi}^{\rm sp}\left( q \right)
  &=&\left[ \hat{I}-\hat{\chi}^{0}\left( q \right) \hat{U}^{\rm sp} \right]^{-1}
   \hat{\chi}^{0}\left( q \right),
  \label{chisp-tpsc} 
  \\
 \hat{\chi}^{\rm ch}\left( q \right)
  &=&\left[ \hat{I}+\hat{\chi}^{0}\left( q \right) \hat{U}^{\rm ch} \right]^{-1}
   \hat{\chi}^{0}\left( q \right), 
  \label{chich-tpsc}
\end{eqnarray}
where $\hat{U}^{\rm sp}$ ($\hat{U}^{\rm ch}$) 
is the local spin (charge) vertex and $\hat{I}$ is the unit matrix. 
The local vertices are determined 
by satisfying two sum rules for the local moment such as 
\begin{eqnarray}
 \frac{2T}{N_c}\sum_{q} \chi^{\rm sp}_{\alpha \alpha} \left( q \right) 
 &=& n_{\alpha}-2\left< n_{\alpha \uparrow} n_{\alpha \downarrow} \right>, 
 \label{chisp}
\\%
 \frac{2T}{N_c}\sum_{q} \chi^{\rm ch}_{\alpha \alpha} \left( q \right)
 &=& n_{\alpha}+2\left< n_{\alpha \uparrow} n_{\alpha \downarrow} \right>
 -n_{\alpha}^{2}, 
 \label{chisp}
\end{eqnarray}
where $n_{\alpha}$ is the particle number at the site $\alpha$. 
We have used the relations 
$n_{\alpha \uparrow} = n_{\alpha \downarrow} = n/2$ 
and $n_{\alpha \sigma} = n_{\alpha \sigma}^{2}$ from the Pauli principles.

The local spin vertex $\hat{U}^{\rm sp}$ is related with 
the double occupancy 
$\left< n_{\alpha \uparrow} n_{\alpha \downarrow} \right>$ 
by the following ansatz 
\begin{eqnarray}
 U^{\rm sp}_{\alpha \alpha} 
 =\frac{ \left< n_{\alpha \uparrow} n_{\alpha \downarrow} \right> }
 { \left< n_{\alpha \uparrow} \right> \left< n_{\alpha \downarrow} \right> }
 U^{0}_{\alpha \alpha}, 
 \label{Usp-tpsc}
\end{eqnarray}
where $U^{0}_{\alpha \alpha}$ is the ($\alpha$ $\alpha$)-element of 
the on-site interaction matrix $\hat{U}^{0}$. 
Equation (\ref{Usp-tpsc}) breaks 
the particle-hole symmetry and should be used for $n_{\alpha} \le 1$.  
When $n_{\alpha} > 1$, that can be applied 
through the particle-hole transformation, 
then the double occupancy 
$D_{\alpha} = \left< n_{\alpha \uparrow} n_{\alpha \downarrow} \right> $ 
is given by 
\begin{eqnarray}
D_{\alpha}
=\frac{U^{\rm sp}_{\alpha \alpha} }{ U^{0}_{\alpha \alpha} }
\frac{n_{\alpha}^2 }{ 4 }
+\left( 1-\frac{U^{\rm sp}_{\alpha \alpha} }{ U^{0}_{\alpha \alpha} } \right)
\left( n_{\alpha}-1 \right)
\theta \left( n_{\alpha}-1 \right), 
\end{eqnarray} 
where $\theta \left( x \right)$ is Heaviside step function. 
Equations (\ref{chisp-tpsc})-(\ref{Usp-tpsc}) give 
a set of the self-consistent equations for the TPSC method. 
Obtaining the $\hat{U}_{\rm sp}$ and $\hat{U}_{\rm ch}$, 
the interaction for the self-energy is obtained as 
\begin{eqnarray}
 \hat{V}^{\Sigma}\left( q \right) = 
 \frac{1}{2} 
 \left[ \hat{U}^{\rm sp} \hat{\chi}^{\rm sp}\left( q \right) \hat{U}^{\rm 0}  
+\hat{U}^{\rm ch} \hat{\chi}^{\rm ch}\left( q \right) \hat{U}^{\rm 0}
 \right]. 
 \label{int_selfene}
\end{eqnarray}
Using the eq. (\ref{int_selfene}), 
the self-energy is given by 
\begin{eqnarray}
 \Sigma_{\alpha \beta} \left( k \right)
 = \frac{T}{N_c}
 \sum_{q} V^{\Sigma}_{\alpha \beta}\left( q \right) 
 G_{\alpha \beta}\left( k-q \right), 
 \label{selfene}
\end{eqnarray}
and the dressed Green's function is obtained as 
\begin{eqnarray}
 \hat{G}\left( k \right) &=& \hat{G}^{0}\left( k \right)
  +\hat{G}^{0}\left( k \right) 
  \hat{\Sigma} \left( k \right) \hat{G}\left( k \right). 
 \label{gr}
\end{eqnarray}
Since we need two sites per unit cell, 
$\hat{U}^{0}$, $\hat{U}^{\rm sp}$, $\hat{U}^{\rm ch}$, 
$\hat{\chi}^{0}\left( q \right)$, 
$\hat{\chi}^{\rm sp}\left( q \right)$, 
$\hat{\chi}^{\rm ch}\left( q \right)$, 
$\hat{V}^{\Sigma}\left( q \right)$, 
$\hat{\Sigma}\left( k \right)$, 
$\hat{G}^{0}\left( k \right)$ and 
$\hat{G}\left( k \right)$ 
all become 2$\times$2 matrices. 
In the present study, 
the spin susceptibility is obtained as the larger eigenvalue of 
the 2$\times$2 spin  susceptibility matrix. 
We consider not only the spin susceptibility, but  
also other physical values such as 
the local spin vertex and the double occupancy. 
In the present calculation, 
we take the system size as $64 \times 64$ $k$-meshes 
and $16384$ Matsubara frequencies.

\section{Results} 
\label{Results} 

\subsection{first principles band calculation} 
\label{first principles band calculation}

\begin{figure}[!htb]
 \centering
  \includegraphics[width=8.0cm]{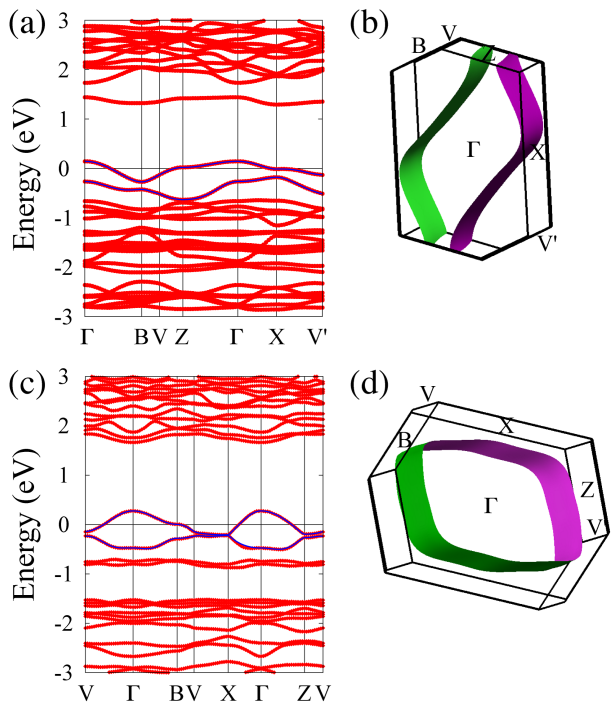}
 \caption{ (color online) 
 (a) Calculated first principles band structure and 
 (b) Fermi surface for the I$_{3}$ salt, 
 (c) first principles band and 
 (d) Fermi surface for the SbF$_{6}$ salt. 
 In both figures of the band structures, 
 the red curves represent the first principles bands  
 and the blue solid curves gives the tight-binding fit.
 }
 \label{fig2}
\end{figure}

Figures \ref{fig2} (a) and (c) show the first principles band structures 
for the I$_{3}$ and the SbF$_{6}$ salts. 
For both materials, the experimental lattice structure 
at an ambient pressure and room temperature are used. 
In both of the materials, it can be seen that 
the highest-occupied molecular orbital (HOMO) 
is isolated from 
the lowest-unoccupied molecular orbital (LUMO). 
Considering this and also 
the number of donor molecules in a unit cell, 
we adopt the HOMO and HOMO$-$1 bands as the target bands 
to construct an effective tight-binding model. 

Although the difference is only the anion, 
the band structures of the two materials are apparently very different. 
In order to reveal the origin of this difference in the band structure, 
in the following we focus on the following two differences of the two salts. 
One is the magnitude of the molecular dimerization, 
namely the dimerization of the donor molecule in the I$_{3}$ salt 
is larger than that in the SbF$_{6}$ salt 
resulting in a larger gap between HOMO and HOMO$-$1 in the former. 
The other is the anisotropy of the band structure, 
namely, there are two flat portions near the Fermi level 
around the Z and the X-points in the I$_{3}$ salt, 
while there is only one flat portion around the B-point in the SbF$_{6}$ salt.

Figure \ref{fig2} (b) shows the Fermi surface of 
the first principles band calculation for the I$_{3}$ salt,  
where the high symmetry points in the Brillouin zone are presented only on 
the $k_{\rm Y} (k_{b})=0$ plane. 
The Fermi surface of the I$_{3}$ salt is disconnected, 
namely quasi-one-dimensional, but  
it is actually close to two dimensional because a slight shift of the 
band structure around the Z-point would give a closed (i.e. 2D) Fermi surface.
Figure \ref{fig2} (d) shows the Fermi surface 
of the SbF$_{6}$ salt. 
The Fermi surface is cylindrical, 
reflecting the two-dimensionality of this salt 
as shown in our previous work 
\cite{Aizawa-Kuroki-NJP-14-113045}.

\subsection{Effective tight-binding model}
\label{Effective model}

\begin{figure}[!htb]
 \centering
  \includegraphics[width=8.0cm]{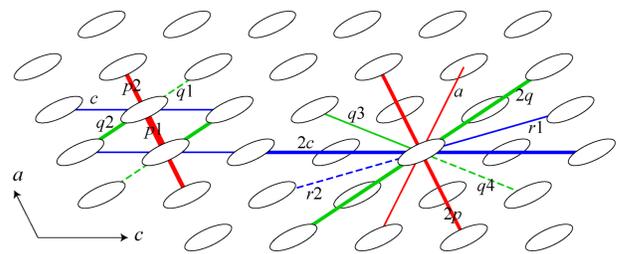}
 \caption{ (color online) 
The tight-binding model for $\beta$-(BDA-TTP)$_{2}X$, 
where left (right) panel shows 
the first (second) nearest-neighbor transfer energies. }
\label{fig3}
\end{figure}

Figure \ref{fig3} shows the effective tight-binding model 
adopted to fit the first principles band. 
The nearest-neighbor transfers are 
shown in the left panel of Fig. \ref{fig3}, and in addition 
we also need to introduce the next-nearest-neighbor transfers 
shown in the right panel of Fig. \ref{fig3}
to reproduce the first principles band structure more accurately.
Note that the stacking direction of the BDA-TTP molecules is 
taken in the $a$-direction\cite{note_coodinates}. 
The band dispersions of the tight-binding model 
are shown as blue solid curves 
in Fig. \ref{fig2} (a) for the I$_{3}$ salt 
and Fig. \ref{fig2} (c) for the SbF$_{6}$ salt. 

The transfer energies for the two salts are summarized 
in Table \ref{tab1}. 
The bottom three lines represent 
the magnitude of the dimerization which is measured by the ratio $t_{p2}/t_{p1}$, 
and the transfer between the inter-stacking direction 
normalized by the average value of intra-stacking transfer energies, 
$(t_{q1}+t_{q2})/(t_{p1}+t_{p2})$ and $2 t_{c}/(t_{p1}+t_{p2})$.
From Table \ref{tab1}, 
it can be seen that there are two major differences between the two salts. 
One is the strength of the molecular dimerization, 
namely the dimerization in the I$_3$ salt is larger than that in 
the SbF$_6$ salt. 
Another difference is the transfers in the inter-stacking direction 
namely, 
the magnitudes as well as the sign of the inter-stacking transfers 
are different between the two salts, that is in $c$($q$)-direction 
in the I$_3$ (SbF$_6$) salt. 
\begin{table}[!htb]
\centering
 \caption{List of the transfer energies in the unit of eV 
 for $\beta$-(BDA-TTP)$_{2}X$.} 
 \begin{tabular}{ c|c|c }
 \hline
  \hspace{20pt} $X$ \hspace{20pt} & 
  \hspace{20pt} I$_{3}$ \hspace{20pt} &
  \hspace{20pt} SbF$_{6}$ \hspace{20pt}   \\ 
 \hline \hline
  $t_{p1}$          (eV)  &           -0.174 &           -0.153 \\
  $t_{p2}$ \phantom{(eV)} &           -0.102 &           -0.126 \\
  $t_{q1}$ \phantom{(eV)} & \phantom{-}0.018 &           -0.071 \\
  $t_{q2}$ \phantom{(eV)} & \phantom{-}0.041 &           -0.055 \\
  $t_{c }$ \phantom{(eV)} & \phantom{-}0.062 & \phantom{-}0.007 \\
  $t_{2c}$ \phantom{(eV)} & \phantom{-}0.002 & \phantom{-}0.005 \\
  $t_{2p}$ \phantom{(eV)} & \phantom{-}0.006 & \phantom{-}0.021 \\
  $t_{a }$ \phantom{(eV)} &           -0.001 & \phantom{-}0.003 \\
  $t_{2q}$ \phantom{(eV)} & \phantom{-}0.004 & \phantom{-}0.005 \\
  $t_{q3}$ \phantom{(eV)} &           -0.012 & \phantom{-}0.003 \\
  $t_{q4}$ \phantom{(eV)} & \phantom{-}0.013 & \phantom{-}0.006 \\
  $t_{r1}$ \phantom{(eV)} & \phantom{-}0.002 & \phantom{-}0.014 \\
  $t_{r2}$ \phantom{(eV)} & \phantom{-}0.009 & \phantom{-}0.008 \\ 
 \hline \hline
  $t_{p2}/t_{p1}$         & \phantom{-}0.586 & \phantom{-}0.824 \\ 
  $\frac{t_{q1}+t_{q2}}{t_{p1}+t_{p2}}$  
                          &           -0.214 & \phantom{-}0.452 \\ 
  $\frac{2 t_{c}}{t_{p1}+t_{p2}}$  
                          &           -0.449 &           -0.050 \\ 
 \hline
 \end{tabular}
 \label{tab1}
\end{table}

\begin{figure}[!htb]
 \centering
  \includegraphics[width=8.0cm]{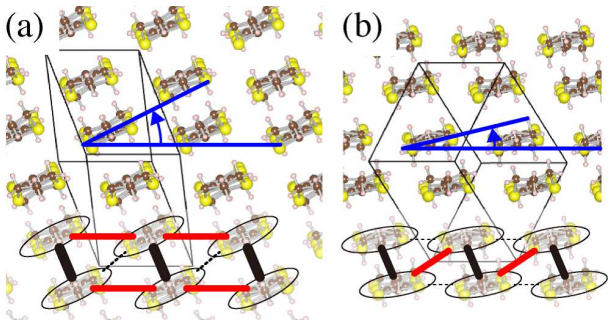}
 \caption{ (color online) 
The lattice structures of (a) the I$_{3}$ and (b) the SbF$_{6}$ salt. 
Blue solid lines represent the tilting angles of the donor molecules, 
which is measured from the $c$-direction taken in the horizontal direction. 
In the lower panel, 
the ellipses represent the donor molecules, 
the black solid lines show the intra-stacking transfer, and 
the red solid (black dotted) lines represent 
the main (not main) inter-stacking transfer. }
 \label{fig4}
\end{figure}

To clarify the origin of the differences between the two salts, 
we consider the alignments of the donor molecules 
in the conducting $c$-$a$ plane for the two salts. 
The conducting $c$-$a$ plane for each salt is shown in Fig. \ref{fig4}. 
We find that 
the tilting angle of the donor molecules from the $c$-axis 
is different between the two salts. 
In the I$_{3}$ salts shown in Fig. \ref{fig4} (a), 
the tilting angle is larger than that in the SbF$_{6}$ salts 
shown in Fig. \ref{fig4} (b). 
The difference in the tilting angle gives rise to 
differences in both the magnitude and the sign
of the main inter-stacking transfers, 
which is $t_{c}$ in the I$_{3}$ salt 
shown in the lower panel of Fig. \ref{fig4} (a), 
while they are $t_{q1}$ and $t_{q2}$ in the SbF$_{6}$ salt
shown in the lower panel of Fig. \ref{fig4} (b). 

Now, let us try to decompose these differences. 
We consider a case where we hypothetically eliminate the dimerization effect. 
Namely, we simplify the model by considering only 
the nearest neighbor transfer energies, 
and replace the hopping in the $p$- and $q$-directions 
by taking their averages. 
The band structure of the simplified model is given by 
\begin{eqnarray}
 \varepsilon \left(\textbf{\textit{k}} \right)
 &=& 2 t_{c} \cos \left( k_{c} \right) + 2 t_{p} \cos \left( k_{a} \right)
 + 2 t_{q} \cos \left( k_{c} + k_{a} \right), 
 \label{simple-band}
\end{eqnarray}
where the transfer energies are 
$t_{p}=-0.138$ eV, $t_{q}= 0.030$ eV, $t_{c}= 0.062$ eV 
for the I$_{3}$ salt, and 
$t_{p}=-0.140$ eV, $t_{q}=-0.063$ eV, $t_{c}= 0.007$ eV 
for the SbF$_{6}$ salt. 
Eliminating the dimerization effect enables us 
to take the unit cell reduced along the $a$-direction. 
By comparing the band structure of the simplified model, 
we can extract the difference caused by the inter-stacking transfer. 
\begin{figure}[!htb]
 \centering
  \includegraphics[width=8.0cm]{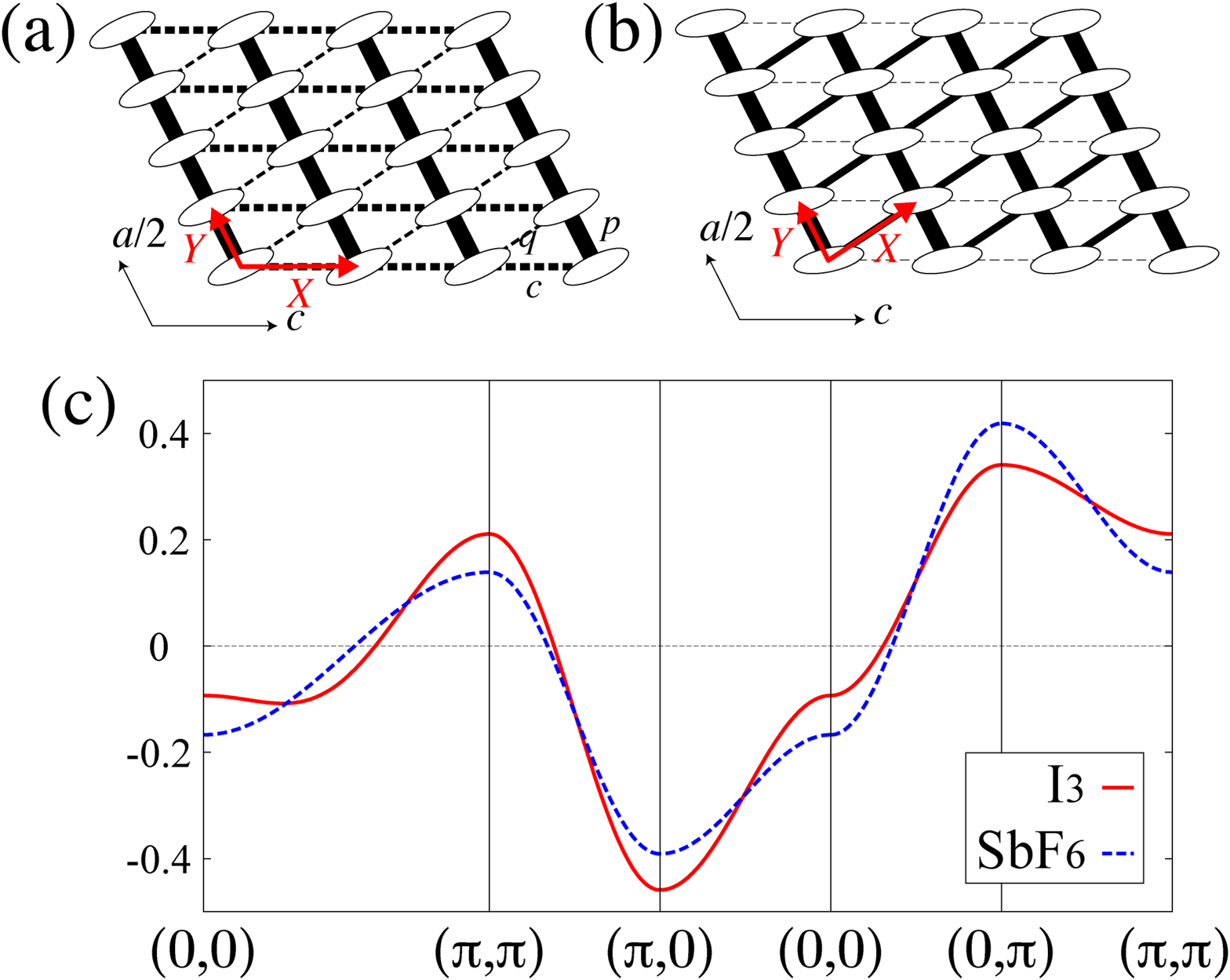}
 \caption{ (color online) 
The simplified model for (a) the I$_3$ salt and (b) the SbF$_6$ salt, 
where the line width schematically represents 
the magnitude of the transfer energies, 
and the solid (dashed) line represents the negative (positive) value 
of the transfer energies. 
(c) The band structure of the simplified model 
by eliminating the dimerization for each salts. }
 \label{fig5}
\end{figure}

We compare the band structure of the two salts in the $(k_X, k_Y)$ plane, 
where $k_Y$ is taken in the molecular stacking direction and 
$k_X$ is taken in the direction of the main inter-stacking transfer, 
namely, $c$-direction in the I$_3$ salt (Fig. \ref{fig5} (a)) 
and $a/2+c$-direction in the SbF$_6$ salt as seen in Fig. \ref{fig5} (b). 
Also, we shift the wave-number by $(\pi, 0)$ for the SbF$_6$ salt 
considering the sign difference in  
the main transfer energies along the inter-stacking direction. 
By such a transformation, 
we find that the band structures between the I$_3$ salt 
and the modified SbF$_6$ salt become very similar 
as shown in Fig.  \ref{fig5} (c). 
Since the simplified model eliminates the dimerization effect, 
the difference in the original band structure between the two salts 
\textit{comes mainly from the dimerization}, 
and the differences coming from the inter-stacking transfer are not essential.

\subsection{Effect of electron correlation cooperating with dimerization}
\label{Effect of dimerization}

A quarter-filled system effectively 
becomes a half-filled system by increasing the 
dimerization\cite{Kino-Fukuyama-JPSJ-65-2158}, 
so that the electron correlation is strengthened. 
Since we now know that the strength of the dimerization is 
the essential difference between the I$_3$ and SbF$_6$ salts,  
we expect that 
the difference of the ground state 
physical properties between the two salts 
is caused by the strength of the electron correlation originating from 
the difference in the strength of the dimerization. 

The strength of the electron correlation can be 
measured by calculating the spin susceptibility. 
We apply the TPSC scheme to the Hubbard model of the I$_3$ salt. 
From the first principles calculation of the I$_3$ salt, 
the band width $W$ is about 0.77eV, 
so we take the on-site interaction $U_{0}=0.8$eV as same as the band width. 
The bare on-site interaction $U_0$ is estimated 
in the other strongly correlated organic conductors 
applying the first-principles calculation 
\cite{Nakamura-Yoshimoto-JPSJ-78-083710,Nakamura-Yoshimoto-PRB-86-205117}. 
Referring to them, the on-site interaction we taken is appropriate.

\begin{figure}[!htb]
 \centering
  \includegraphics[width=8.5cm]{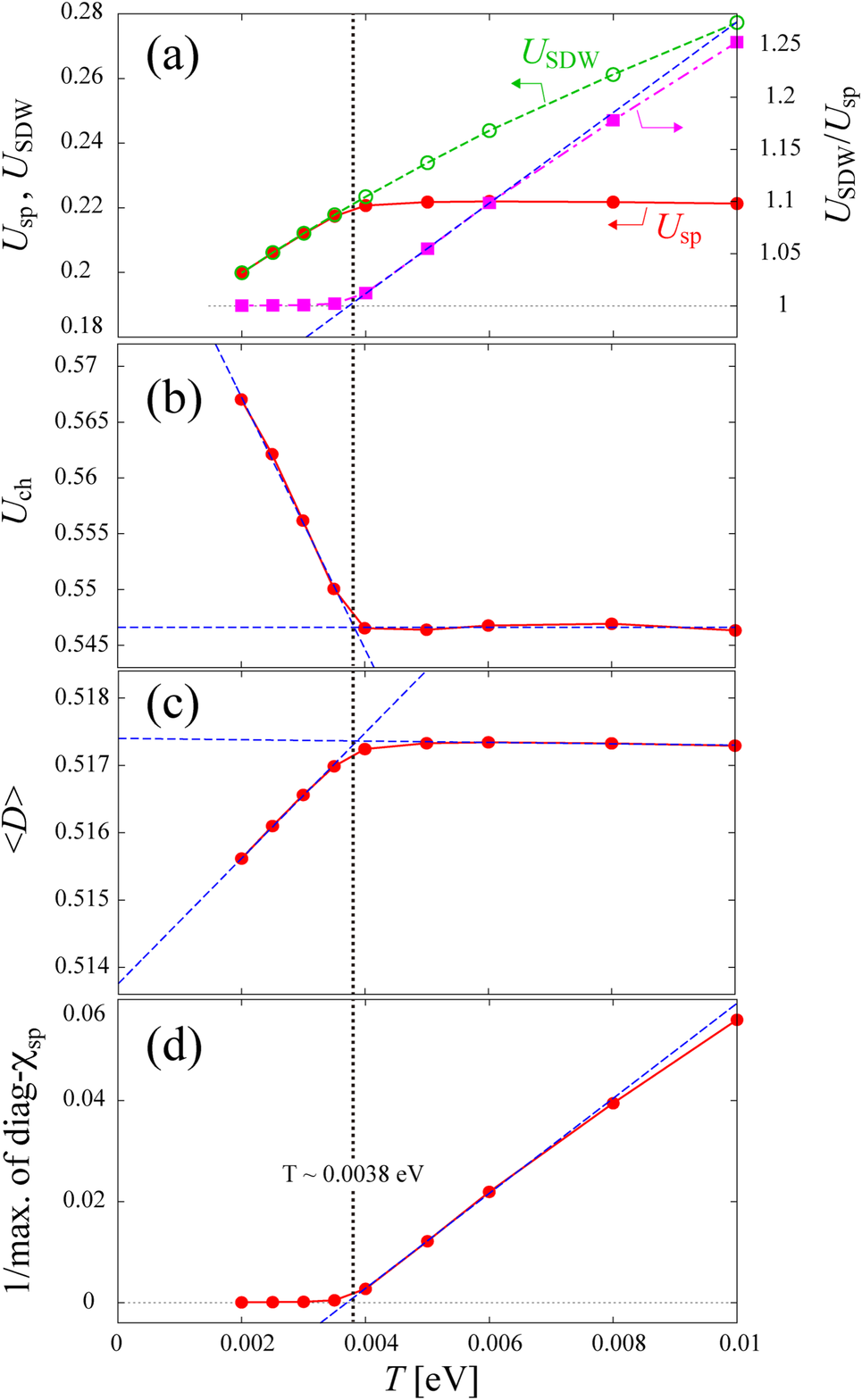}
 \caption{(color online) 
Temperature dependence of 
(a) $U_{\rm sp}$ and $U_{\rm SDW}$ for left scale and 
$U_{\rm SDQ}/U_{\rm sp}$ for right scale, 
(b) $U_{\rm ch}$, 
(c) double occupancy $\left< D \right>$, and 
(d) maximum value of the diagonalized spin susceptibility 
for the model of the I$_3$ salt. 
Blue dashed lines represents the line extrapolating the each values and 
black dotted lines are about $T = 0.0038$ eV. 
}
 \label{fig6}
\end{figure}

Figure \ref{fig6} (a) shows 
the temperature dependence of the local vertex of the spin part $U_{\rm sp}$ 
and the critical on-site interaction of the magnetic order $U_{\rm SDW}$ 
in the left scale. 
Above the temperature $T \approx 0.004$eV, 
$U_{\rm sp}$ is almost unchanged and 
$U_{\rm SDW}$ gradually decreases with lowering the temperature.
Below $T \approx 0.004$eV, 
$U_{\rm sp}$ takes almost the same value, 
but somewhat smaller value than $U_{\rm SDW}$, 
which can be understood that the magnetic ordering 
is developed with lowering the temperature.

In the right scale of Fig.\ref{fig6}(a), 
we present the ratio $U_{\rm SDW}/U_{\rm sp}$ as a function of $T$. 
The TPSC approach satisfies the Mermin-Wagner theorem 
so the true magnetic ordering does not occur in the present model, 
but we can regard the temperature at which the line extrapolating 
$U_{\rm SDW}/U_{\rm sp}$ from high temperature reaches unity 
as the magnetic critical temperature in the actual three dimensional system.  
We estimate the magnetic critical temperature to be about 0.0038 eV. 
Reflecting the tendency toward the magnetic ordering, 
$U_{\rm ch}$ quickly increases below $T = 0.0038$ eV 
as shown in Fig. \ref{fig6} (b). 
We show the double occupancy 
$\left< D \right>=\left< n_{\uparrow}n_{\downarrow} \right>$ 
as a function of $T$ in Fig. \ref{fig6} (c). 
Similarly to the local vertices, 
the double occupancy $\left< D \right>$ also changes below $T = 0.0038$eV. 
Decreasing the temperature reduces the double occupancy, 
which means the tendency of the magnetic localization at each site. 
Figure \ref{fig6} (d) shows 
the inverse of the maximum value of the spin susceptibility against $T$. 
As expected from Fig. \ref{fig6} (a), 
the inverse of the spin susceptibility extrapolates 
to zero around $T = 0.0038$ eV. 
In fact, a very recent experiment observes a magnetic transition 
in the Mott insulating state of the $I_3$ salt at low temperature
\cite{Isono-private-communication}. 
TPSC is not capable of directly describing the magnetic 
ordering of a Mott insulator, but the very fact that the material is a 
Mott insulator is consistent with our view that the electron correlation 
effect is strong  due to the strong dimerization.

\begin{figure}[!htb]
 \centering
  \includegraphics[width=7.0cm]{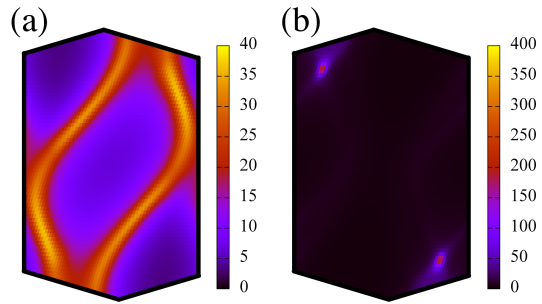}
 \caption{(color online) 
(a) The absolute value of the Green's function and 
(b) the diagonalized spin susceptibility 
for the model of the I$_3$ salt at $T=0.004$eV. 
}
 \label{fig7}
\end{figure}

Figures \ref{fig7} (a) and (b) show 
the absolute value of the Green's function $\left| G \right|$ 
and the spin susceptibility $\chi_{\rm sp}$ of the I$_3$ salt 
with $U_0=0.8$eV and $T=0.004$eV. 
The absolute value of the Green's function takes large values near 
the Fermi surface shown in Fig. \ref{fig7} (a). 
The wave number at which the spin susceptibility is maximized 
corresponds to the nesting vector of the Fermi surface 
as seen in Fig. \ref{fig7} (b). 
As shown in Fig. \ref{fig7} (b), 
the maximum value of the spin susceptibility takes a large value 
since its temperature is close to the critical temperature. 

To clarify the relation between the electron correlation and the dimerization, 
we measure the strength of the dimerization by the quantity $t_{p2}/t_{p1}$. 
When $t_{p2}/t_{p1}$ goes to unity, the dimerization decreases. 
If the decrease of the dimerization results in 
weakening the electron correlation, 
we expect 
(i) $U_{\rm sp}$ gradually deviates from $U_{\rm SDW}$, 
(ii) the double occupancy $\left< D \right>$ becomes large, 
and (iii) the maximum value of the spin susceptibility decreases 
within the TPSC scheme. 
Furthermore, 
if the stronger electron correlation of the I$_3$ salt originates from the 
stronger molecular dimerization, 
all the quantities should approach the values close to those 
of the SbF$_6$ salt when the dimerization is reduced hypothetically 
in the model of the I$_3$ salt.

\begin{figure}[!htb]
 \centering
  \includegraphics[width=7.0cm]{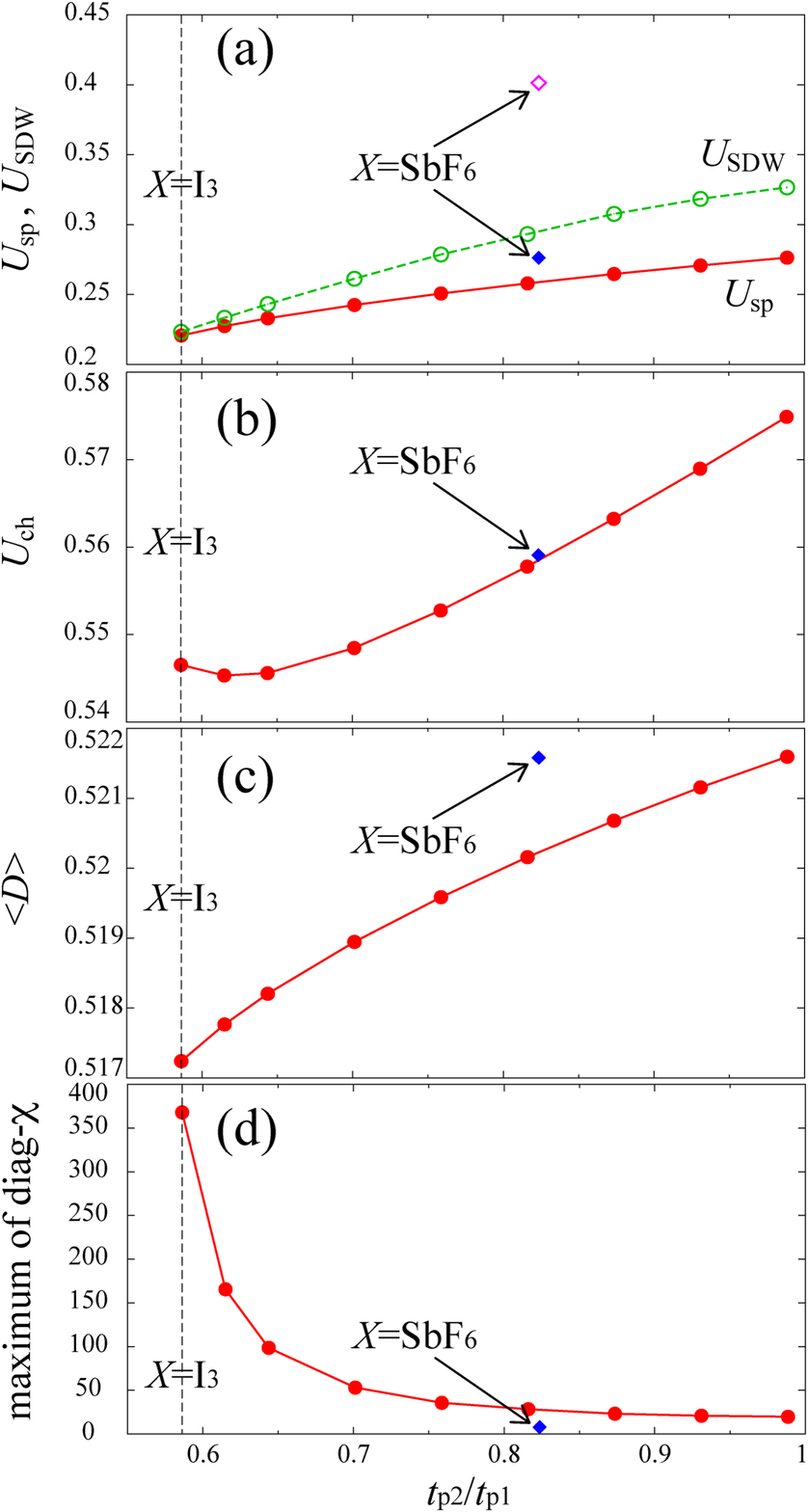}
 \caption{(color online) 
The strength of the dimerization $t_{p2}/t_{p1}$ dependence of 
(a) $U_{\rm sp}$ and $U_{\rm SDW}$, (b) $U_{\rm ch}$, 
(c) double occupancy $\left< D \right>$, and 
(d) maximum value of the diagonalized spin susceptibility 
for the model of the I$_3$ salt with $U_0=0.8$eV and $T=0.004$eV. 
The each values on the SbF$_6$ salt with same $U$ and $T$ 
are also shown in the corresponding figures. 
}
 \label{fig8}
\end{figure}

Let us now investigate the relation between 
the electron correlation and the dimerization. 
Figure \ref{fig8} (a) shows 
the local vertex of the spin part $U_{\rm sp}$ and 
the critical on-site interaction for the magnetic order $U_{\rm SDW}$ 
as a function of $t_{p2}/t_{p1}$ in the model of the I$_3$ salt, 
also shows them for the SbF$_6$ salt at the point corresponding $t_{p2}/t_{p1}$, 
where we take $T=0.004$ eV. 
Decreasing the dimerization (increasing $t_{p2}/t_{p1}$), 
$U_{\rm sp}$ gradually differs from $U_{\rm SDW}$, 
which expects that increasing the $t_{p2}/t_{p1}$ suppresses 
the maximum value of the spin susceptibility. 
In contrast to the temperature dependence, 
decreasing the dimerization increases $U_{\rm ch}$ 
as seen in Fig. \ref{fig8} (b), 
although $U_{\rm sp}$ differs from the $U_{\rm SDW}$. 
In Fig. \ref{fig8} (c), 
the double occupancy $\left< D \right>$ monotonically increases 
with decreasing the dimerization, 
which can be understood as the suppression of the magnetic localization. 
This tendency is confirmed by the 
deviation of $U_{\rm sp}$ from $U_{\rm SDW}$. 
Figure \ref{fig8} (d) shows 
the maximum value of the spin susceptibility 
as a function of $t_{p2}/t_{p1}$. 
Decreasing the dimerization from the actual value of the I$_3$ salt 
quickly suppresses the spin susceptibility, 
and that of the I$_3$ salt takes almost the same value as that of 
the SbF$_6$ salt around the same strength of the dimerization. 

From the $t_{p2}/t_{p1}$ dependence in Fig. \ref{fig8}, 
we can say that the electron correlation in the I$_3$ salt 
is stronger than in the SbF$_6$ salt due to the strong dimerization. 
We therefore conclude that 
the difference of the ground state between the two salts, namely, 
insulating for the I$_3$ salt and superconducting for the SbF$_6$ salt, 
originates from the strength of the dimerization, 
which affects the electron correlation. 
Applying the pressure to the I$_3$ salt reduces the dimerization, 
resulting in the metallicity, and hence the superconductivity appears. 

\section{Conclusion}
\label{Conclusion}

In the present study, 
we have performed first principles band calculations  
and have derived the effective tight-binding models 
of $\beta$-(BDA-TTP)$_{2}$I$_{3}$ and $\beta$-(BDA-TTP)$_{2}$SbF$_{6}$. 
The band structures and the Fermi surface between the I$_3$ and SbF$_6$ salts 
are apparently different although only the anion differs. 
The derived tight-binding models, 
which accurately reproduce the first principles band structures 
of the two salts, 
show that the differences between the two salts 
comes mainly from the strength of the dimerization.

As for the effect of the electron correlation, 
we have presented the TPSC results for quantities 
such as the spin susceptibility in the Hubbard model for the two salts. 
The TPSC results show that 
the electron correlation becomes stronger upon lowering the temperature 
and/or increasing the dimerization strength. 
Then, we have hypothetically reduced 
the strength of the dimerization in the I$_3$ salt to that of the SbF$_6$ salt, 
where all the calculated quantities tend to become similar to 
those of the SbF$_6$ salt. 
Thus, we conclude that the electron correlation in the I$_3$ salt 
is stronger than the SbF$_6$ salt due to the strong dimerization. 
The expected stronger correlation in the I$_3$ salt is at least 
qualitatively consistent with a recent experimental observation 
that the material is a Mott insulator, which is a hallmark of strong 
correlation, and exhibits a magnetic transition at low temperature
\cite{Isono-private-communication}. 
Applying the pressure to the I$_3$ salt reduces the dimerization, 
which weakens the electron correlation, 
and hence the superconductivity appears as in the SbF$_6$ salt. 

In the present study, we have considered only the on-site (intra-molecular) 
electron-electron interaction. It remains an interesting future problem 
to study the effect of the off-site interactions.
In fact, it has been known that in organic conductors 
having quarter-filled bands,  
the Mott insulating state often competes with the charge ordering and/or 
charge-density-wave states\cite{Seo-Merino-JPSJ-75-051009}. 
It is an interesting issue to investigate how such interactions would affect 
the insulating properties as well as the mechanism of the superconductivity.

\section*{Acknowledgment}
We thank T. Isono for showing the latest experimental data.
This work is supported by Grant-in-Aid for Scientific Research from 
the Ministry of Education, Culture, Sports, Science and Technology of 
Japan, and from the Japan Society for the Promotion of Science. 
Part of the calculation has been performed at the 
facilities of the Supercomputer Center, 
ISSP, University of Tokyo.



\begin{thebibliography}{99}
\bibitem{huckel-comment}
Recently, several studies report that 
the first-principles band structure slightly differs from 
those obtained by the extended H$\ddot{\rm u}$ckel method
\cite{Kandpal-Opahle-PRL-103-067004,Nagai-Nakamura-PRB-83-104523,Alemany-Pouget-PRB-89-155124}.
\end{thebibliography}
\end{document}